# A Simple Model to Measure Linewidth Enhancement Factor (α) for Multi-wavelengths Semiconductor Laser

Samir K Mondal
Central Scientific Instruments Organisation (Council of Scientific and Industrial Research, India), Chandigarh, India
E-mail: samirmondal01@gmail.com

**Abstract:** A novel and simple model is proposed for the measurement of the linewidth enhancement factor (LEF) α of multiwavelengths semiconductor laser. It is based on the suppression characteristics of an arbitrary mode while other modes are optically injection locked individually. The proposed model can be used to measure α value of individual modes of multi-wavelengths laser with large mode spacing. As a proof of the concept, the model is used to experimentally measure α of a mode of a slotted Fabry- Pérot laser source.

## I. INTRODUCTION

The linewidth enhancement factor (α) is one of the most critical parameters for semiconductor laser and it controls the laser dynamics, especially for widely tunable lasers. It is important to estimate value of α for a laser source to design high speed photonics circuit. The LEF is responsible for excess line broadening, and chirp under high speed modulation and performance degradation. The LEF arises as the result of an asymmetric gain profile in semiconductor laser. The LEF for a semiconductor laser is usually encountered in the range of 2-10. There are several methods for LEF measurement and each has its own merits and demerits. Hakki-Paoli method [1] is the most common one to measure α. Other methods are optical feedback self-mixing [2], different approaches under optical injection locking technique [3,4]. The injection locking technique exploits the asymmetric locking width under external optical injection when α≠0 caused by asymmetric gain profile.

This work proposes a novel and simple model for measuring value of α of a multiwavelengths Fabry-Pérot semiconductor laser using boundary values of external optical injection locking width. The model can be used for multichannel laser source with large mode spacing, where Hakki-Paoli method may not be applied. The model is experimentally verified on a slotted Fabry-Pérot laser. The experiment uses a slotted multiwavelengths Fabry-Pérot (SFP) laser source having 7 modes with mode spacing of ~8 nm [5]. Although it is single-mode at wavelength λ=1536 nm in free running condition, the laser source can work in any of other 6 modes individually under guided optical injection locking condition. Under optical injection locking all modes get suppressed gradually except the locked mode and the suppression is guided by the injection locking condition of the locked mode and its α value. The suppression characteristic is used to find value of α.

## II. MODEL

The analytical model exploits the boundary values of locking width $\Delta\omega_{(-\rho\sqrt{1+\alpha^2} \le \Delta\omega \le \rho(\theta = \frac{\pi}{2}))}$ [6,7] of a mode under optical injection resulting in suppression of other modes. The locking width Δω (=$\omega_{inj}$-$\omega_l$) is the difference between injected angular frequency ($\omega_{inj}$) and angular frequency ($\omega_l$) of the locked laser mode respectively. The suppression of other modes are determined by α value of the locked mode under optical injection. The α value can be extracted from the typical suppression characteristic of the suppressed modes. For the laser mode *l*1 of a multichannel laser source, the steady state photon flux for the mode becomes [8]

$$P_{l1} = R_{sp}(1 + G_{l1} + G_{l1}^{\ 2} + .......)$$
$$= \frac{R_{sp}}{(1 - G_{l1})} \quad (G_{l1} \le 1 \text{ below threshold}) \qquad (1)$$

where $G_{l1}$ and $R_{sp}$ are the round trip loop gain, rate of spontaneous radiation respectively. Similarly for an adjacent side-mode *l*2, the photon flux is given by

$$P_{l2} = R_{sp}(1 + G_{l2} + G_{l2}^{\ 2} + .......)$$
$$= \frac{R_{sp}}{(1 - G_{l2})} \qquad (2)$$

$G_{l2}$ being the round trip loop gain of the mode.
Considering *l*2 becomes active under optical injection locking, the side-mode suppression (SMS) can be accordingly expressed as [8]

$$SMS = \frac{P_{l2}}{P_{l1}} = \frac{(1 - G_{l1})}{(1 - G_{l2})} \qquad (3)$$

When a mode is guided to radiate under external optical injection, the gain of the respective mode is presented by [9]

$$G_{l2} = \frac{1}{\tau} - 2\rho\cos\theta \qquad (4)$$

Where $\rho = k_c\sqrt{\frac{I_{inj}}{I_{mod}}}$ is coupling factor between injected ($I_{inj}$) and the respective laser mode ($I_{mode}$) power and θ is phase angle between injected field and respective laser mode field.

The phase angle θ is given by [10]

$$\theta = \cos^{-1}\left(\frac{\Delta\omega}{\rho\sqrt{1+\alpha^2}}\right) - \tan^{-1}\left(\frac{1}{\alpha}\right) \qquad (5a)$$

$$\Delta\omega = \rho(\sin\theta - \alpha.\cos\theta) \qquad (5b)$$

where Δω (=$\omega_{inj}$–$\omega_l$) is the locking width $(-\rho\sqrt{1+\alpha^2} \le \Delta\omega \le \rho(\theta = \frac{\pi}{2}))$ and α is linewidth enhancement factor.

Optical injection to any of the modes suppresses the main (most prominent) mode (*l*1) gradually with the detuning of the injected wavelength. The suppression and its dependence on to the injected detuning frequency can be expressed as

$$SMS = \frac{(1-G_{l1})}{(1-\frac{1}{\tau}+2\rho\cos\theta)}$$

and

$$\frac{\partial(SMS)}{\partial\omega_{inj}} = -\frac{(1-G_{l1})}{(1-\frac{1}{\tau}+2\rho\cos\theta)^2}\cdot\left(-2\cdot\rho\cdot\sin\theta\frac{\partial\theta}{\partial\omega_{inj}}\right)$$

$$= \frac{2(1-G_{l1})}{(1-\frac{1}{\tau}+2\rho\cos\theta)^2}\cdot\frac{\sin\theta}{\cos\theta+\alpha.\sin\theta} \qquad (6)$$

respectively.

It is well known that the injection locking width is guided by boundary value of Δω given by

$$\Delta\omega_{boundary} = \begin{cases} \theta = \pi/2 & \text{..................} \quad 7(a) \\ -\rho\sqrt{1+\alpha^2} & \text{..................} \quad 7(b) \end{cases}$$

From the above expression and Equ.5, the boundary values of θ are obtained as

$$(a) \to \theta = \frac{\pi}{2} \qquad (8)$$

and

$$(b) \to \theta = \cos^{-1}\left(\frac{\Delta\omega}{\rho\sqrt{1+\alpha^2}}\right) - \tan^{-1}\left(\frac{1}{\alpha}\right)$$
$$= \cos^{-1}(-1) - \tan^{-1}\left(\frac{1}{\alpha}\right)$$
$$= \pi - \tan^{-1}\left(\frac{1}{\alpha}\right) \qquad (9)$$

Replacing value of θ in Eqn.6 we obtain following slopes for positive (short wavelength) and negative (longer wavelength) detuning sides as

$$\frac{\partial(SMS)}{\partial\omega_{inj}} = \frac{2(1-G_{l1})}{(1-\frac{1}{\tau})\alpha} \qquad (10a)$$

and

$$\frac{\partial(SMS)}{\partial\omega_{inj}} = \frac{2(1-G_{l1})}{(1-\frac{1}{\tau}+2\rho\cos\theta)}\cdot\tan\theta\frac{1}{1+\alpha.\tan\theta}$$
$$= \frac{2(1-G_{l1})}{(1-\frac{1}{\tau}+2\rho\cos\theta)}.\left(-\frac{1}{\alpha}\right)\frac{1}{1+\alpha(-\frac{1}{\alpha})}$$
$$\approx \infty \to \frac{\pi}{2}(slope) \qquad (10b)$$

respectively.

As one sees in the slope equation, the shorter wavelength side is function of 'α'and $G_{l1}$, which can be used to find α value. A proof of concept experiment is performed to find 'α' value of a mode in a multichannel SFP laser source using the model.

## III. EXPERIMENTAL SETUP & PROCEDURE

The SFP used in the experiment has eight slots towards the rear facet of the cavity. The design details of the source can be found in author's work, Ref.5. The wavelength-locking experimental setup is shown in Fig. 1. [5]

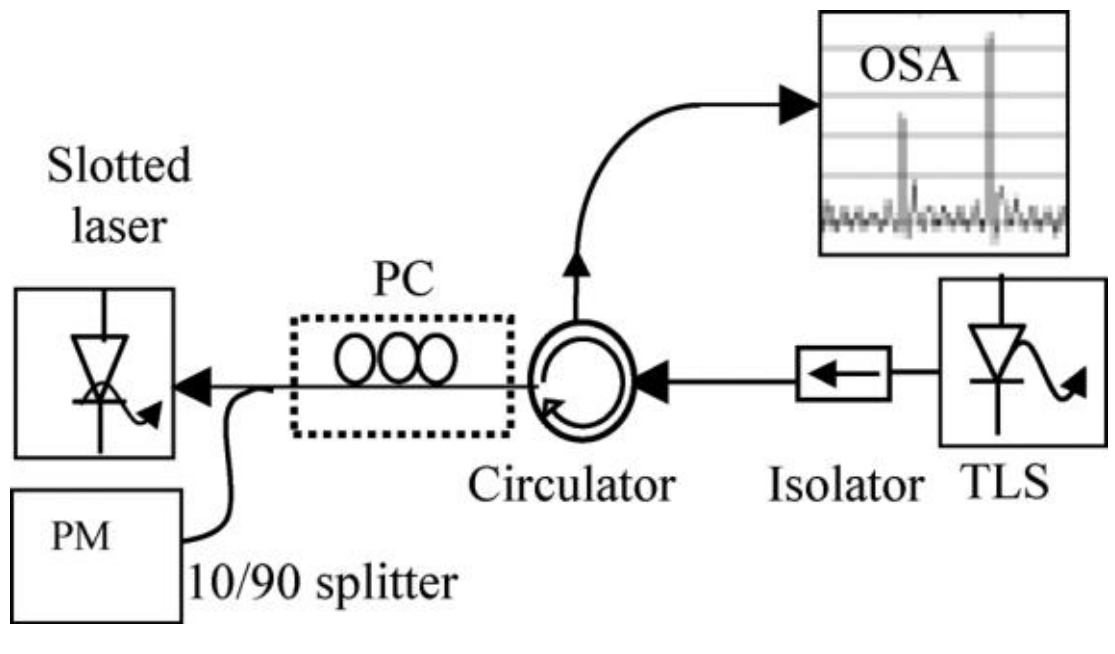

Fig.1 Experimental setup

An ANDO tunable laser source is used as the master laser (ML) that injects optical power to the SFP laser through an optical isolator, circulator, and power splitter. The polarization controllers control the injected field to match the SFP output polarization and maximize the response. The output from the SFP is directed to an optical spectrum analyzer (OSA) through the circulator for spectral analysis of the emission.

The SFP has structurally resonating wavelengths between 1510 and 1565 nm determined by the free-spectral range $\Delta\lambda = \left(\frac{\lambda^2}{2nL_s}\right)(\approx 8nm)$ of the SFP subcavity. There is no designed pre-selection of which mode will radiate and the lasing mode is determined by the optical path to the facets. However, it radiates single-mode (*l*1) at1536.82 nm.

For our experiment, the external temperature of the SFP is held at $23^0$C and the current is held at threshold. The mode (*l*1) becomes the most prominent mode. The mode (*l*2) at 1528 nm is chosen for optical injection locking.

## III. RESULTS & DISCUSSIONS

Fig. 2 presents the free-running spectra of the laser source at threshold current.

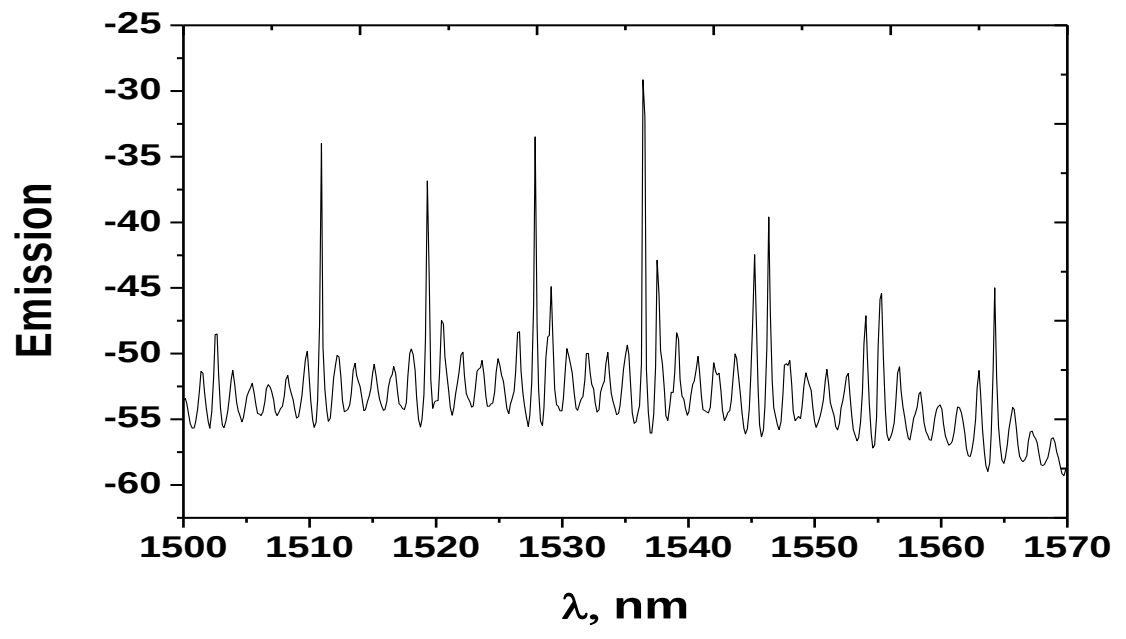

Fig.2 Free running spectrum of the multichannel source at current of I=43 mA

Fig. 3 plots the recorded SMS on the OSA as a function of the injection wavelength for an injection power of -20 dBm. The injecting wavelength is tuned in the step of 0.01 nm. The nature of suppression of main mode varies with the tuning of injected wavelength. The positive and negative detuning may be understood from rate equation analysis with detuning boundary given by $(-\rho\sqrt{1+\alpha^2} \le \Delta\omega \le \rho(\theta = \frac{\pi}{2}))$. A closer look at the plot in Fig.3 (a) shows gradual changes in the modal power suppression from the beginning of the injection

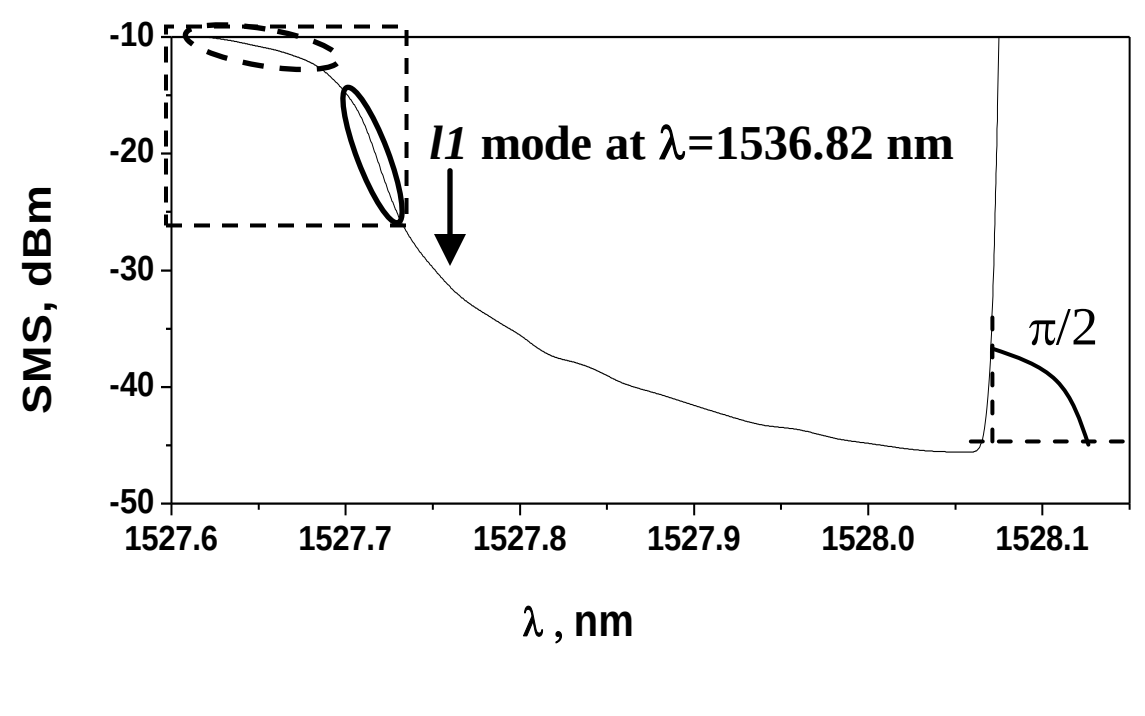

(a)

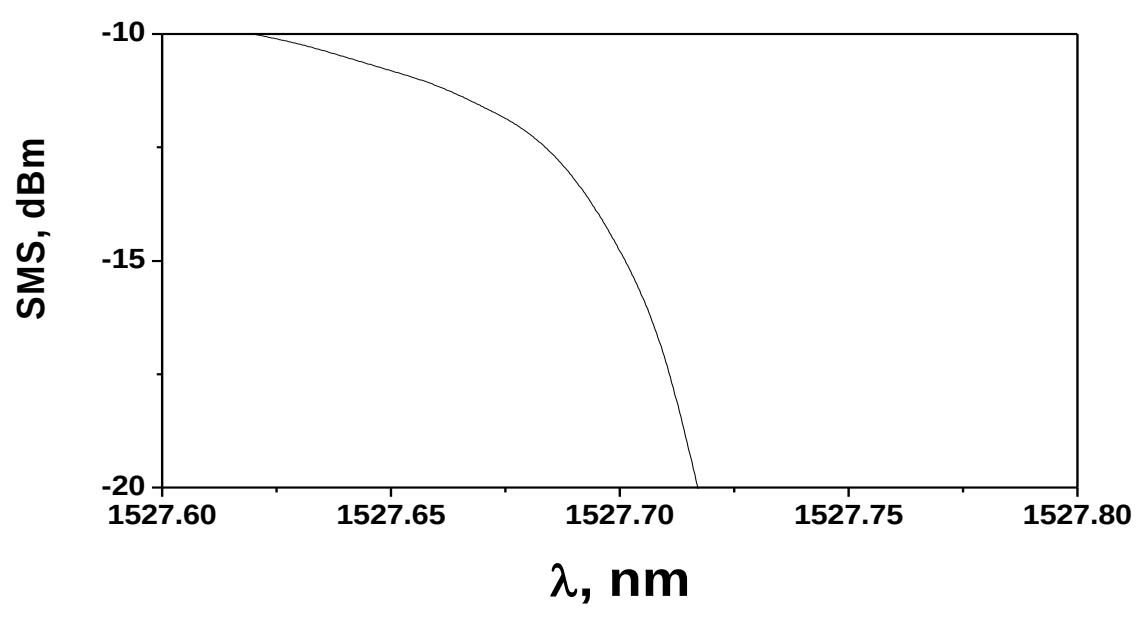

(b)

Fig.3 (a) plots suppression of main mode at λ=1536.82 nm as the mode at λ=2528 nm is locked by optical injection locking and de-tuning, (b) represents the selected part of Fig. 3(a), which is used to find ‘α’.

locking (positive detuning boundary) boundary to the end of the locking (negative detuning boundary) boundary. The typical suppression nature can be described by Equ.10. From the slope Equ.10(b) one can find that the slope of the locking boundary toward negative detuning boundary (condition 7b) tends to π/2, which is also observed experimentally as shown in Fig.3 (a). Toward the positive detuning boundary (condition 7a) it should follow Equ.10a, which is function of α. The Table 1 presents the value of α obtained by using experimental data and the model.

Table 1 showing calculated value of ‘α’ of *l*2 mode using consecutive slopes, Fg. 3(a), with approximated $G_{l1}$. 3(b), $(\frac{1}{\tau} << 1)$.

| ΔSMS | $\Delta f_{\text{inj}}$ ($\Delta\lambda_{\text{inj}}$) | $\alpha = 2(1-G_{l1})\cdot\frac{2\pi\Delta f_{inj}}{\Delta SMS}$ ($\Delta\omega_{inj} = 2\pi\Delta f_{inj}$) |
|---|---|---|
| -5dBm [(-15dBm)-(-10dBm)] $\frac{P_{l2}}{P_{l1}} = 0.3$ | 10.45 GHz (0.08 nm) | 4.38($G_{l1}$-0.99) |
| -5dBm [(-20dBm)-(-15dBm)] $\frac{P_{l2}}{P_{l1}} = 0.3$ | 2.61 GHz (0.02 nm) | 4.37($G_{l1}$-0.96) |

Unlike negative detuning boundary the slope toward positive detuning changes with injected wavelength. The slope in the beginning is marked by ovals of dashed and solid line respectively. The region under dashed oval shape gives value of α=4.38, row 1 in Table 1. It is to be mentioned that the source is excited at close to threshold current. As the mode *l2* is picked up by optical injection locking, *l*1 mode is pushed below threshold due to suppression and accordingly value of $G_{l1}$ is assumed ~ 0.99. However, as one moves to the slope within oval with solid line, the *l*1 mode is pushed below threshold and value of G*l*1is reduced further. The value of $G_{l1}$ is taken as 0.96 in the next slope regime. This region gives α=4.37 for a suppression of -5dBm, row 2, Table 1. The accuracy of α value depends upon $G_{l1}$. If one goes further down to the SMS curve the value of $G_{l1}$ cannot be predicted as the SMS curve changes continuously and hence cannot be used to find α. Though the value of α depends upon value of $G_{l1}$, the approximation of $G_{l1}$ (0.99, 0.96) should be valid as the laser works at threshold current ($G_{l1}$=1) before injection locking. However, the model is applicable if the injected power is sufficient so that the locking width reaches to its maximum value.

## IV. CONCLUSIONS

In conclusion, a novel and simple model is proposed for the measurement of the linewidth enhancement factor α of multi-channel semiconductor laser with large mode spacing. It is based on the suppression characteristics of an arbitrary mode while other modes are optically injection locked individually. The injected optical field should be sufficient so that the locking width reaches to its maximum value. The proposed model is used to experimentally measure α for an arbitrary mode of a slotted Fabry-Pérot laser source. The model may not work for multiwavelengths laser source with small mode spacing (<50GHz). The locking width of individual mode overlaps with next sidemodes with small mode spacing and hence the model fails to measure value of α accurately.

The author acknowledges that the experimental data used in this work obtained from experiments [5] performed by the author during his appointment with Tyndall National Institute, Ireland. The author gratefully acknowledges Brian Corbett for providing the SFP laser source. The author is also grateful to Prof. S.N. Sarkar for fruitful discussion.